\documentclass[twocolumn,showpacs,preprintnumbers,amsmath,amssymb]{revtex4}

\usepackage{graphicx}
\usepackage{dcolumn}
\usepackage{bm}

\begin{document}


\title{Optimal counter-current exchange networks}

\author{Robert S. Farr}
 \affiliation{The London Institute for Mathematical Sciences, 
35a South Street, Mayfair, London, UK, and Unilever R\&D, Colworth Science 
Park, Bedford, MK44 1LQ, UK.}
 \email{robert.farr@unilever.com}

\author{Yong Mao}
 \affiliation{School of Physics and Astronomy, 
University of Nottingham, Nottingham, NG7 2RD, UK}
 \email{yong.mao@nottingham.ac.uk}
\date{\today}

\begin{abstract}
We present a general analysis of exchange devices linking their
efficiency to the geometry of the exchange surface
and supply network.  For certain parameter ranges, we show that the optimal 
exchanger consists of densely packed pipes which can span a thin sheet of
large area (an `active layer'), which may
be crumpled into a fractal surface and supplied with a fractal network
of pipes. We derive the efficiencies of such exchangers,
showing the potential for significant gains compared to regular exchangers
(where the active layer is flat), using parameters 
relevant for biological systems.
\end{abstract}


\maketitle

\section{Introduction}
The design of efficient exchange devices is an important problem
in engineering and biology.
A wide variety of heat exchangers, such as plate, coil and counter-current,
are employed in industrial settings \cite{Perry}, while in nature, 
leaf venation, blood circulation networks, gills and lungs 
have evolved to meet multiple physiological imperatives.
A distinctive feature of the biological examples is their complex,
hierarchical (fractal) nature \cite{West}, with branching and usually 
anastomosing 
geometries \cite{Katifori,Makanya}. It is clear that one reason for
this is the possibility to include a large surface for exchange
within a compact volume, as in the human lungs, which comprise an alveolar
area greater than $50{\rm m}^2$ \cite{Wiebe}. However, maximal surface area
is unlikely to be the only criterion for optimization.
As an example, West {\it et al} have analyzed biological circulatory 
systems on the basis that power is minimized with the constraint that
a minimum flux of respiratory fluid is brought to every cell in the
volume of an organism, and were able to explain well known allometric
scaling laws in biology \cite{West}.

Although scaling behaviors are known in some cases, the detailed geometry
of optimal exchangers remains elusive.
With the advance of new fabrication technologies such as 3D printing 
\cite{Hague}, it is becoming possible to build structures of comparable
complexity to biological systems, so there is a need
not only to understand the principles and compromises
upon which natural systems are based, but also for that understanding to
be constructive, mapping system parameters to actual designs.

The analytic literature in this area has focused on heat transfer from a 
fluid to a solid body, with a particular emphasis on cooling of
integrated circuits \cite{Tuckerman}. Branching fractal networks are much
studied due to their ability to give good heat transfer 
with a low pressure drop \cite{Bejan,Chen} (although sometimes simpler 
geometries can be more efficient \cite{Escher}), and multiscale structures
are also found to have a high heat transfer density \cite{Fautrelle}.

\begin{figure}[b]
\includegraphics[width=1.0\columnwidth]{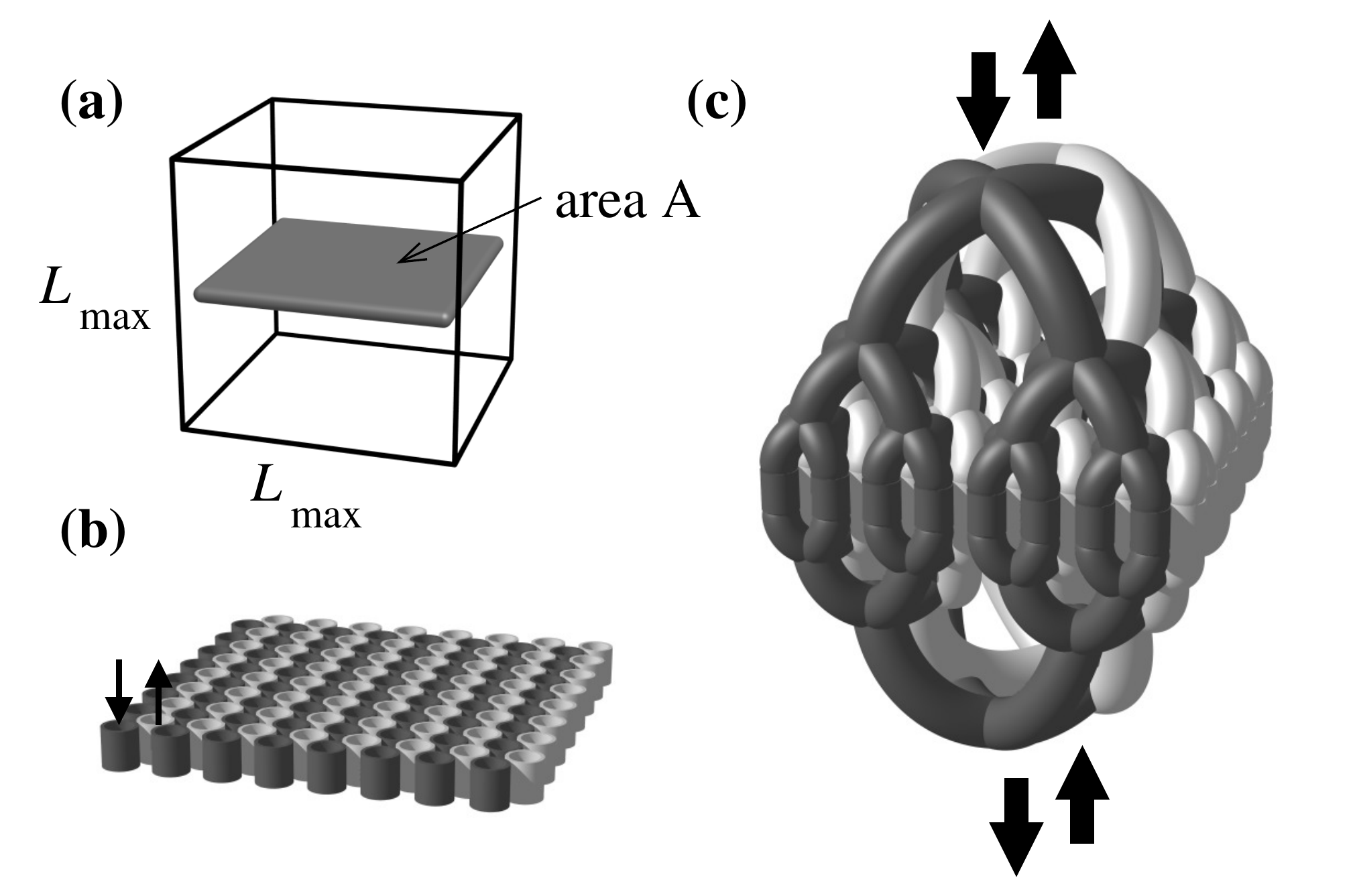}
\caption{\label{surface_and_supply}(a) Schematic of the geometry of a 
counter-current heat exchanger `active layer' fitting inside a
prescribed cubic volume of side length $L_{\max}$.
(b) Detail of the active layer, showing a regular array of 
pipes carrying alternately counter-flowing streams.
(c) The active layer connected to a branching and (on the other side) 
anastomosing fractal supply network.
}
\end{figure}

In this contribution, we consider exchange as a general process, which 
includes gas, solute and heat exchange, and we look for the optimal
designs which can ensure complete exchange (to be defined below)
while requiring a minimum amount of mechanical power to generate
the necessary fluid flows. 

We use the language of thermal processes, since the relevant material 
properties have widely used  notation. However, with a suitable translation 
of quantities, the analysis also applies to mass transfer.
For example, in a thermal system with linear materials, the quantities:
temperature, heat, heat capacity per unit volume and thermal conductivity
would correspond in a system of gas exchange to: partial pressure of gas,
mass of gas, Henry's law coefficient and the product of the Henry's
law coefficient and gas diffusivity. For mass exchange with solutes,
the analogue of temperature would be osmotic pressure of the solute.

\section{Non-dimensionalization}
The first step is to gather problem parameters into dimensionless groups, 
which span the space of possible exchange problems.

Suppose there are two counter-flowing (perhaps dissimilar)
fluids with given properties: thermal conductivities 
$\kappa_j$ ($j\in\{1,2\}$),
heat capacities per unit volume $C_j$ and
viscosities $\eta_j$. Let there be an imposed difference $\Delta T$ 
in the {\em inlet} temperatures, and an imposed volumetric flow rate $Q_1$ of
fluid 1 (while we are free to choose $Q_2$). 
For example, if we are considering thermoelectric generation from the
exhaust gases of a vehicle, $Q_1$ would be the volumetric flow of exhaust 
gases. Analogously, in gas exchange for vertebrate respiration, we take
the required blood flow to the lungs or gills as the fixed quantity $Q_1$.

The fluid streams between which exchange occurs are assumed separated by 
walls of thickness $w$ (taken to be the minimum
consistent with biological or engineering constraints) and
thermal conductivity $\kappa_{\rm wall}$; the latter again an imposed 
constraint. We assume that the exchanger
needs to be compact, in that it fits inside a roughly cubical
volume of side length $L_{\rm max}$, and the pipes, being straight,
are each of length $L\le L_{\max}$.
Last, we wish the exchange process to go to completion, in that
the total exchanged power is of order $E_{\rm end}=C_1 Q_1 \Delta T$, which
results in the outlet temperature of flow 1 being equal to the inlet
temperature of flow 2, and conversely.
Our aim is to find an exchange network which satisfies all these 
constraints (which we believe are a typical set for both engineering and 
biological systems), while requiring the minimum amount of power to drive the 
flow through the network.

To proceed, we non-dimensionalize on $L_{\max}$ and $\kappa_{\rm wall}$,
defining the new quantities:
\begin{eqnarray}
\hat{w} \equiv w/L_{\max},\ \ \ 
\hat{r}_j \equiv r_j/L_{\rm max},\ \ \ 
\hat{L} \equiv L/L_{\max},
\nonumber \\
\hat{A} \equiv A/L_{\max}^{2}\ \ \ {\rm and}\ \ \ 
\hat{\kappa}_j \equiv \kappa_j/\kappa_{\rm wall}.\nonumber
\end{eqnarray}
The specification of the problem can be conveniently reduced to three
non-dimensional parameters, the first two of which capture the 
asymmetry of the two fluids:
\begin{equation}
\beta \equiv (C_1/C_2)^2 (\eta_2/\eta_1)\ \ \ {\rm and}\ \ \ 
\gamma \equiv \kappa_1/\kappa_2 .
\end{equation}
We then note that if all the available volume were filled with
pipes of the smallest possible radius, and the two fluids were set to
uniform temperatures differing by $\Delta T$, then there would be a
maximum possible exchanged power of order
$E_{\rm max}=\Delta T \kappa_{\rm wall} L_{\rm max}^3 /w^2$.
Thus our last parameter is the ratio of the required exchange rate to 
this maximum:
\begin{equation}
\epsilon\equiv E_{\rm end}/E_{\rm max}
=Q_1 C_1 w^2/(L_{\max}^3 \kappa_{\rm wall}), \label{epsilon}
\end{equation}
and we typically expect $\epsilon\ll 1$.

\section{Optimal regular exchangers} 
We consider a regular array of counter-flowing streams in
$N_j$ straight pipes of radii $r_j$ ($j=1,2$)
and length $L$ (the same for both types),
where we initially ignore any feed network to supply the individual pipes.
This regular array is shown in figure \ref{surface_and_supply}(b),
and we describe this array of pipes as the `active layer', since it is
where exchange actually occurs.

To proceed, we make three geometric approximations: First,
assuming roughly circular pipes, we approximate the total cross section 
(perpendicular to flow) of the array as
\begin{equation}
A\approx \pi N_1 (r_1+w/2)^2+\pi N_2 (r_2+w/2)^2.
\end{equation}
Second, let $\alpha$ be the
area across which exchange occurs, then if no clustering of one type occurs
$\alpha$ will be approximately the minimum of the two pipe perimeters,
multiplied by $L$. We thus propose a simple approximation to
the total area across which exchange occurs:
\begin{equation}\label{exarea}
\alpha\approx \left[ (N_1 2\pi r_1 L)^{-1}
+(N_2 2\pi r_2 L)^{-1}\right]^{-1}.
\end{equation} 
Third, we approximate the thermal conductance per unit area across which
exchange occurs to be
\begin{equation}
s \approx \left[
(w/\kappa_{\rm wall})+(r_1/\kappa_1)+(r_2/\kappa_2)\right]^{-1}. \label{s}
\end{equation}

When is exchange complete?  We assume the pipes are slender, so that heat 
diffusion along the length of a pipe
is negligible compared to across its width (and also to advective
transport along its length); and that the temperature
over a cross section perpendicular to its length is roughly uniform.
Let $z$ be the distance along a pipe, with $z=0$ being the upstream end 
of fluid 1 and the downstream end of fluid 2, so the average temperatures 
over cross sections
of each of the two types of pipe are $T_j(z)$. 
We define the difference of inlet temperatures to be
$\Delta T\equiv T_1(0)-T_2(L)$. By considering
the total heat flux per unit length $J(z)$ between the two sets of pipes,
we can write down the material derivative of temperature as each fluid
moves along its respective pipe:
\begin{equation}
\pi N_j r_j^2 C_j \frac{DT_j}{Dt} = (-)^{j}J(z), \label{simul}
\end{equation}
where, since the average flow speed in the pipes of type $j\in\{1,2\}$ is
$Q_j/(N_j \pi r_j^2)$, the material derivative is
\begin{equation}
\frac{D}{Dt} \equiv \frac{\partial}{\partial t}+
(-)^{j+1}\frac{Q_j}{N_j \pi r_j^2}\frac{\partial}{\partial z}.
\end{equation}
If $s$ is the thermal conductance per unit area between pipes we note:
\begin{equation}
J(z) \approx \alpha s [T_1(z)-T_2(z)]/L .
\end{equation}

In the steady state regime, $\partial/\partial t\equiv 0$ so
Eqs.\ (\ref{simul}) lead to an exchanged power $E$ given by
\begin{eqnarray}
\frac{E}{s \alpha\Delta T} &=& 
\frac{\xi_1 \xi_2(e^{1/\xi_1}-e^{1/\xi_2})}{\xi_2 e^{1/\xi_1}-\xi_1 e^{1/\xi_2}}
\approx \min(1,\xi_1,\xi_2) \label{E_P}\\
\xi_j&\equiv& Q_j C_j/(\alpha s). \label{xi}
\end{eqnarray}
Complete exchange means $E\approx C_1 Q_1 \Delta T$, which from 
Eq.\ (\ref{E_P}) is equivalent to $\xi_1 \le \xi_2$ and $\xi_1\le 1$. 
We note from the analysis accompanying Eq.\ (\ref{E_P}) that there is a
special case of a `balanced' exchanger, in which $Q_1 C_1 = Q_2 C_2$ 
(so $\xi_1 = \xi_2$) and the change of temperature
with $z$ for both streams is linear, rather than being exponential.
The optimal exchanger should have this property, since otherwise
some of the pipe length will contribute to dissipated power but not exchange. 
Thus $Q_2$ is determined by the imposed value of $Q_1$.

\begin{table}
\begin{tabular}{l|lll}
\hline\hline
System: & T.E.G.\hspace{2em} & Pigeon\hspace{2em} & Salmon \\
\hline
Exchanged: & Heat & Oxygen & Oxygen \\
\hline
$L_{\max}/{\rm m}$       & $2.0\ 10^{-1}$ & $5.0\ 10^{-2}$ & $2.0\ 10^{-2}$ \\
$w$/m                    & $5.0\ 10^{-4}$ & $5.0\ 10^{-7}$ & $5.0\ 10^{-7}$ \\
$Q_1/$m$^{3}$s$^{-1}$    & $5.0\ 10^{-2}$ & $2.0\ 10^{-5}$ & $1.0\ 10^{-6}$ \\
$C_1$/S.I.               & $1.0\ 10^{3} $ & $2.0\ 10^{-6}$ & $2.0\ 10^{-6}$ \\
$C_2$/S.I.               & $1.0\ 10^{3} $ & $1.3\ 10^{-5}$ & $1.0\ 10^{-7}$ \\
$\kappa_1$/S.I.          & $4.0\ 10^{-2}$ & $1.8\ 10^{-16}$ & $1.6\ 10^{-16}$ \\
$\kappa_2$/S.I.          & $4.0\ 10^{-2}$ & $2.3\ 10^{-10}$ & $1.6\ 10^{-16}$ \\
$\kappa_{\rm wall}$/S.I. & $1.0\ 10^{1} $ & $1.8\ 10^{-16}$ & $1.6\ 10^{-16}$ \\
$\eta_1$/Pa\,s           & $4.0\ 10^{-5}$ & $4.0\ 10^{-3}$ & $4.0\ 10^{-3}$ \\
$\eta_2$/Pa\,s           & $4.0\ 10^{-5}$ & $4.0\ 10^{-5}$ & $1.0\ 10^{-3}$ \\
\hline
$\beta$                  & $1.0\ 10^{0}$ & $2.4\ 10^{-4}$ & $1.0\ 10^{2}$ \\
$\gamma$                 & $1.0\ 10^{0}$ & $7.8\ 10^{-7}$ & $1.0\ 10^{0}$ \\
$\epsilon$               & $1.6\ 10^{-4}$ & $4.4\ 10^{-4}$ & $3.9\ 10^{-4}$ \\
\hline
$r_{1,\rm reg}$/m        & $1.0\ 10^{-3}$ & $2.5\ 10^{-5}$ & $5.2\ 10^{-6}$ \\
$r_{2,\rm reg}$/m        & $1.0\ 10^{-3}$ & $2.2\ 10^{-6}$ & $2.1\ 10^{-5}$ \\
$A_{\rm reg}$/m$^2$      & $4.0\ 10^{-2}$ & $2.5\ 10^{-3}$ & $4.0\ 10^{-4}$ \\
$L_{\rm reg}$/m          & $2.0\ 10^{-1}$ & $5.0\ 10^{-2}$ & $2.0\ 10^{-2}$ \\
$P_{\rm reg}$/W          & $2.4\ 10^{1}$ & $6.2\ 10^{-1}$ & $7.7\ 10^{-1}$ \\
\hline
$r_{1,\rm frac}$/m       & $5.1\ 10^{-4}$ & $5.0\ 10^{-6}$ & $5.0\ 10^{-6}$ \\
$r_{2,\rm frac}$/m       & $5.1\ 10^{-4}$ & $5.4\ 10^{-7}$ & $7.3\ 10^{-6}$ \\
$A_{\rm frac}$/m$^2$     & $6.6\ 10^{-2}$ & $1.0\ 10^{-2}$ & $7.6\ 10^{-4}$ \\
$L_{\rm frac}$/m         & $4.3\ 10^{-2}$ & $7.1\ 10^{-4}$ & $2.9\ 10^{-3}$ \\
$P_{\rm frac}$/W         & $1.8\ 10^{1}$ & $6.0\ 10^{-2}$ & $4.0\ 10^{-1}$ \\
\hline\hline
\end{tabular}
\caption{\label{reality}Estimated parameters for various real systems. 
`S.I.' refers to the international system of units; so for thermal systems
$C$ will have units ${\rm J m}^{-3}{\rm K}^{-1}$ and $\kappa$
will have units ${\rm W m}^{-1}{\rm K}^{-1}$. For gas exchange,
$C$ will have units kilogram of relevant gas per m$^{3}$ of fluid,
per Pascal of partial pressure, and $\kappa$ will have units 
${\rm kg\,s}^{-1}{\rm m}^{-1}{\rm Pa}^{-1}$ (so that $\kappa/C$ is
a diffusivity).
`T.E.G.' is thermo-electric generation from internal combustion engine exhaust
(we have chosen values corresponding to a car/personal automobile).
For the animal respiratory systems we assume that transport across
the exchange membrane is similar to that of water. For blood, we assume that 
oxygen can exist in a mobile form (dissolved in the water-like serum) and
an immobile form (bound to haemoglobin). Thus the oxygen `conductivity' 
$\kappa_1$ for blood is the
same as for water, while $C_1$ is increased over that of water 
by the carrying capacity of haem. 
Data are from Refs.\ \cite{CRC,Cussler,Schmidt,Butler,Kicenuik}. 
Results for a regular exchange network are indicated by the subscript
`reg'; while the results for the
fractal exchange surfaces (denoted by a subcript `frac')
use a Hausdorff dimension $d=2.33$.
For the cases of pigeon and salmon respiration, we impose the additional
constraint that $r_1>5\mu\rm m$, in order to allow erythrocytes to
pass through blood vessels (type 1 pipes). 
This appears to only affect the fractal case, and without this 
requirement, the optimized value of $r_1$ for
this fractal case would be $1.5\mu\rm m$ and $0.4\mu\rm m$ 
for pigeon and salmon respectively.
}
\end{table}

\begin{figure}[!t]
\includegraphics[width=0.8\columnwidth]{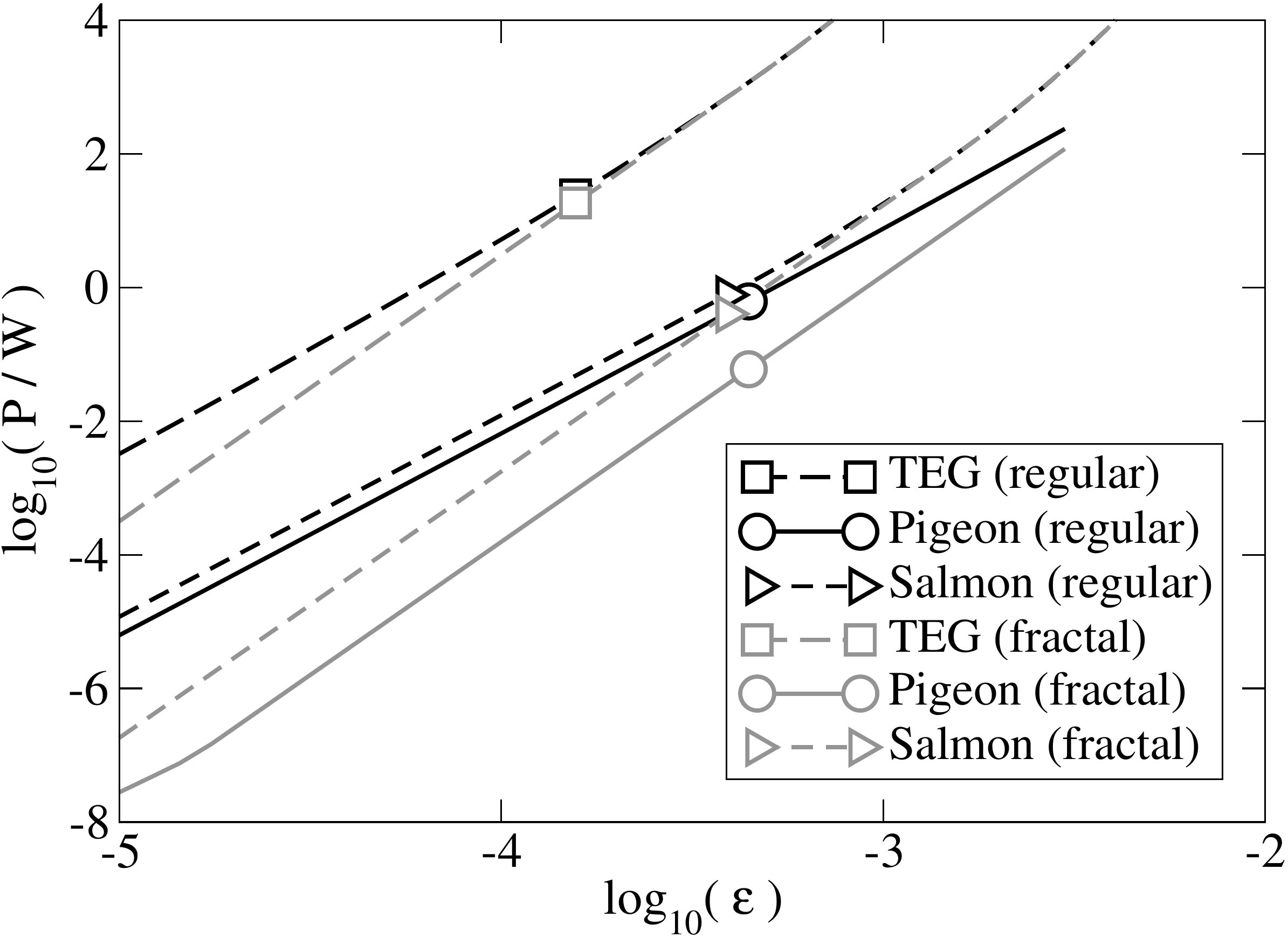}
\caption{\label{P_vs_epsilon} Plots of power dissipated in exchange
for the three cases of table \ref{reality}. Here we change $Q_1$ to achieve 
different values of $\epsilon$. The actual cases in table \ref{reality}
are shown as symbols. For the cases of `pigeon' and `salmon', we
additionally impose the constraint that blood vessels (type 1 pipes)
should be large enough to carry erythrocytes (taken as the condition
$r_1>5\mu\rm m$).
}
\end{figure}

\begin{figure}
\includegraphics[width=1.0\columnwidth]{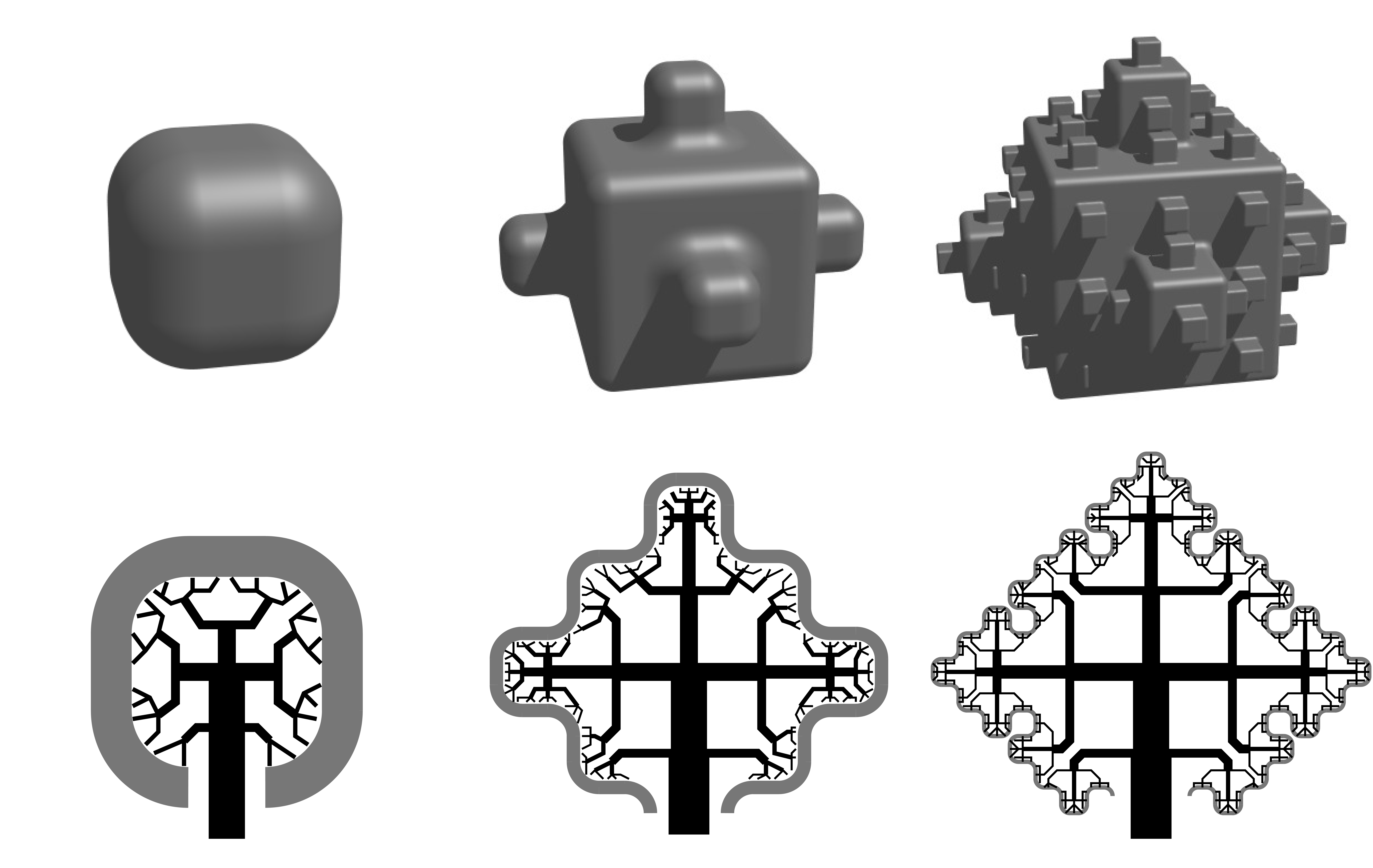}
\caption{\label{fractal_surface}Top row: schematic of the active layer of 
figure \ref{surface_and_supply}(a), corrugated into a hierarchical
(fractal) surface, comprising (left to right) greater area and more
iterations of the fractal. Bottom row: Schematic
section through these surfaces showing the fractal supply network 
in the interior (the corresponding network outside is not shown, and
will require a more complex design to ensure equal flow to all
parts of the active layer).
}
\end{figure}

Now we seek to minimize the total power $P$ required to run the
exchanger, $P=Q_1\Delta p_1+Q_2\Delta p_2$,
where $\Delta p_j$ are the pressures dropped across the two types of pipes.
For laminar (Poiseuille) flow, and using the `balanced'
condition $Q_1 C_1 = Q_2 C_2$ to eliminate $Q_2$, we obtain:
\begin{eqnarray}
P &=& P_0 \epsilon^2 \hat{L}\left(
\frac{1}{N_1 \hat{r}_1^4}+\frac{\beta}{N_2 \hat{r}_2^4}\right) ,
\label{P_nondim}\\
P_0 &\equiv& 8 \eta_1 \kappa_{\rm wall}^2 L_{\max}^3 / 
(\pi w^4 C_1^2 ).
\end{eqnarray}
Our task is to minimize the power $P$ to drive the flow 
in Eq.\ (\ref{P_nondim}) by choosing the five
quantities $N_j$, $\hat{r}_j$ and $\hat{L}$, while ensuring
the exchanger is compact (fits in the required volume):
\begin{eqnarray}
\max(\hat{r}_j)\le\hat{L}\le 1 , \label{Lconstr} \\
\hat{A}=\pi N_1(\hat{r}_1+\hat{w}/2)^2 +
\pi N_2(\hat{r}_2+\hat{w}/2)^2 \le 1 , \label{Aconstr}
\end{eqnarray}
and also that exchange is complete, which from $\xi_1 \le 1$ and 
Eqs.\ (\ref{exarea}), (\ref{s}), (\ref{xi}) leads to 
\begin{eqnarray}
\frac{\epsilon}{\hat{w}^2 2\pi\hat{L}}
\left(\frac{1}{N_1 \hat{r}_1}+\frac{1}{N_2 \hat{r}_2}\right)
\left( \hat{w}+
\frac{\hat{r}_1}{\hat{\kappa}_1} + \frac{\hat{r}_2}{\hat{\kappa}_2}\right)
\le 1. \label{Econstr}
\end{eqnarray}
The optimization can then be performed numerically 
with the constraints (\ref{Lconstr}), (\ref{Aconstr}) and (\ref{Econstr}).
We do this in two different ways, which give essentially identical
results: we either repeatedly choose a random direction in the five
dimensional space of $(N_j,\hat{r}_j,\hat{L})$ and follow this
direction until either the dissipated power does not fall or a constraint
is encountered; or, we impose completeness of exchange in Eq.\ (\ref{Econstr})
as an equality, which allows us to determine $\hat{L}$ given the
other variables, and then perform an exhaustive search for the minimum
power over the more tractable 4-dimensional space $(N_j,\hat{r}_j)$.

Table \ref{reality} shows the geometry of some optimized regular
exchangers for real cases, and the optimized results are included in 
figure \ref{P_vs_epsilon} with the label `regular'.

\section{Scaling of regular exchangers and limiting conditions}
It is interesting to look at what limits the exchanger efficiency in
different cases. For the examples studied here, the numerical results show
that over essentially the
entire range of $\epsilon$, Eqs.\ (\ref{Aconstr}) and, unsurprisingly,
(\ref{Econstr}) are satisfied as equalities. Furthermore, $\hat{w}$
is typically much less than $\hat{r}_j$ or $\hat{r}_j/\hat{\kappa}_j$. 

For symmetric exchangers, where $N_1 = N$, $\hat{r}_1 = \hat{r}_2$ and
$\hat{\kappa}_1 = \hat{\kappa}_2$, we can see the consequences of this 
for the scaling behavior with $\epsilon$, because
Eqs.\ (\ref{Aconstr}) and (\ref{Econstr}) reduce to
\begin{equation}
2 \pi N_1 \hat{r}_1^2 \approx 1 \ \ \ {\rm and} \ \ \ 
\frac{\epsilon}{\hat{w}^2 2\pi \hat{L}}
\cdot
\frac{2}{N_1 \hat{r}_1}
\cdot
\frac{\hat{2r}_1}{\hat{\kappa}_1}
\approx 1,
\end{equation}
which implies the dissipated power 
\begin{equation}
P\approx 16\pi P_0 \epsilon^3 / (\hat{\kappa}_1 \hat{w}^2).
\end{equation}
Two observations follow from this rough analysis: First, the
dissipated power in this approximation does not depend on $\hat{L}$, so
that although the numerical results indicate that optimization
pushes $\hat{L}$ towards unity, this is only a weakly
selected result. Thus exchangers with very similar dissipated power
can be made from rather thin active layers [as shown schematically
in figure \ref{surface_and_supply}(b)] without incurring a strong penalty. 
This is useful in allowing room for the supply network that we will 
wish to attach to the active layer.

Second, an interesting question to ask for an optimal exchange network
is: which constraint is significantly limiting the performance?
The non-trivial constraint in this case is typically the area $\hat{A}$ of
the active layer in Eq.\ (\ref{Aconstr}), which we would prefer to make
larger than $L_{\max}^2$.

Taken together, these observations imply that a route to further optimization
is to have an active layer which is both thin and also folded in some way 
to accommodate a larger area inside the prescribed volume of the device;
an approach we will pursue further in section \ref{Double} below.

\section{The branched supply network}
So far, we have considered the active layer of the exchanger as
an independent entity. However, it must be supplied with the two
working fluids, and for the optimization scheme above to be relevant, this
supply network, which dissipates power but performs no significant exchange, 
should not dominate the power consumption of the whole device.
 
Consider therefore a branched (and fractal) supply network shown in 
figure \ref{surface_and_supply}(c), which brings the streams to the
exchanger's active layer. In contrast to Ref.\ \cite{West},
we do not need the supply network to pass close to every point in space.
Suppose that each pipe comprising the
supply network branches into $b$ smaller pipes at each hierarchical level $k$
of the tree (where pipes with higher values of $k$ are smaller, and closer 
to the active layer where exchange 
occurs). Let the ratio of pipe radii between neighboring
levels be $\rho<1$, and the ratio of pipe lengths be $\lambda$.
The ratio of power dissipated between hierarchical levels is therefore
\begin{equation}
P_{k+1}/P_{k}=\lambda/(b\rho^4).
\end{equation}
Since the active layer is densely covered with pipes, the condition
to fit the supply network into space is $\rho \ge b^{-1/2}$.
Therefore, provided $\lambda>b\rho^4$, the power will increase exponentially 
with $k$ and the overall power dissipation in the supply network 
will be of order that in the last layer; and therefore of the same order
as in the active layer. The supply network will therefore not dominate
the power dissipation.

\section{Fractal exchange networks\label{Double}}
From the solution above for optimum regular exchange networks, the lateral
cross section $A$ always expands to its maximum value $L_{\max}^2$.
If this restriction were lifted, a more efficient
exchanger would likely be possible.
This can be achieved by allowing the active layer (provided it is thin
enough, and can still be provided with a branching supply network) to 
become corrugated, while still fitting within the prescribed roughly
cubical volume $L_{\max}^{3}$ available. One way to do this is to turn the 
active layer into an approximation to a fractal surface.
Thus suppose the active layer to corrugated into such a fractal surface over a 
range of lateral length scales down to a scale $x\ge L$ (where $L$ is the 
pipe length, and therefore the thickness of the layer). In the limit
$x\rightarrow 0$ the surface would have some Hausdorff 
dimension \cite{Hausdorff}, which we denote $d$.
Figure \ref{fractal_surface} shows schematically an example in which the 
surface is the type I quadratic Koch surface with (in the limit) Hausdorff 
dimension $d_{\rm koch}=\ln 13/\ln 3\approx 2.33$. 
Let the area of the active layer be $A(x)$, where 
$A(L_{\rm max})=L_{\rm max}^{2}$, then from Hausdorff's definition of
dimension, we see that $A(x)=L_{\rm max}^{2}(x/L_{\rm max})^{2-d}$.
We can therefore replace the inequality $\hat{A}\le 1$ in 
Eq.\ (\ref{Aconstr}) by
\begin{equation}
\hat{A}=\pi N_1(\hat{r}_1+\hat{w}/2)^2 +
\pi N_2(\hat{r}_2+\hat{w}/2)^2 \le \hat{L}^{2-d}. \label{Afracconstr}
\end{equation}
Figure \ref{P_vs_epsilon} shows the effect of $\epsilon$ (varied through
altering $Q_1$) on the power dissipation for fractal
exchangers corresponding to the scenarios in table \ref{reality},
compared to that of the regular exchanger.
Corrugating the exchange layer into a type I quadratic Koch surface
leads to a significant reduction in the dissipated power for the two
biological cases (factor gain of $10$ for pigeon lungs and $2$ for
salmon gills). 
However, the small size of the optimum pipe radii $r_1$ may mean that 
this degree of optimization is precluded by other considerations.
For instance, erythrocytes need to be able to pass through these
type 1 (blood carrying) vessels.

Crumpling the active layer into a (limited length scale) fractal surface
would also be expected to produce a novel scaling of dissipated
power with $\epsilon$. The numerical results indicate that in the
optimum exchanger $\hat{A}$ expands to
its new maximum extent, so Eq.\ (\ref{Afracconstr}) is an equality.
As above, Eq.\ (\ref{Econstr}) is an equality, but we find for
the TEG case that $\hat{w}$ is comparable to $\hat{r}_{j}$, while
$\hat{w}$ remains substantially less than $\hat{r}_{j}/\hat{\kappa}_j$.
This leads in the symmetric case to the following versions of
Eqs.\ (\ref{Afracconstr}) and (\ref{Econstr}):
\begin{equation}
\frac{9 \pi}{4}N_1 \hat{w}^2 \approx \hat{L}^{2-d} \ \ \ {\rm and} \ \ \ 
\left(\frac{2\hat{r}_1}{\hat{\kappa}_1}\right)
\left(\frac{2}{N_1 \hat{r}_1}\right) \approx
\frac{2\pi \hat{L} \hat{w}^2}{\epsilon},
\label{simple_approx_frac}
\end{equation}
(the first assuming for definiteness $\hat{w}\approx\hat{r}_1$),
which implies the dissipated power is
\begin{equation}
P\approx \left(\frac{9}{2}\right)^{\frac{2}{3-d}}
\frac{\pi P_0}{\hat{w}^2} \hat{\kappa}_1^{\frac{1-d}{3-d}}
\epsilon^{\frac{5-d}{3-d}}.
\end{equation}
For the quadratic Koch surface, this leads to $P\propto \epsilon^{4.01}$,
which is close to the observed exponent in figure \ref{P_vs_epsilon}.

\section{Conclusions}
Exchange networks of the class we show here exhibit broadly power-law
dependence of the dissipated power with the quantity $\epsilon$,
which measures the required throughput: the rate of exchange of
heat, gas or solute needed.  This is true both for a
fractally corrugated or a simple regular array of exchange pipes.
However, the fractal exchangers demonstrate gains in efficiency
when compared to regular exchangers for small values of $\epsilon$,
and in particular for parameters relevant to biological systems.
This is driven by the higher efficiency of a thin active exchange layer
of large area; the fractal corrugations being one way to accommodate 
this geometry in a compact volume.

We note that the 
analysis we have performed here aims specifically to minimize
required power while ensuring complete exchange has taken place and
compactness of the exchange device.
In practice, other design constraints may need to be included, for example
a requirement that the network be robust \cite{Katifori} or easily repairable 
\cite{Qu,Farr} under external attack \cite{Cohen2000,Cohen2001};
or the cost of building the network may be 
significant compared to its operating costs \cite{Bohn,Durand}.
Nevertheless, the conditions analyzed here are, we believe, relevant to
a wide class of engineering and biological systems and could provide
the basis for improved industrial efficiency and insights into the
structures used for respiration in the living world.

\end{document}